\documentclass[sigplan]{acmart}
\usepackage{graphicx} 
\usepackage{tabularx}
\usepackage{array}
\usepackage{fvextra}

\title{Environment-Sensitive Lexicographic Disambiguation for Contextual Parsing}
\date{February 2026}
\author{Alejandro Luis Vaz Mayato}

\begin{document}

\begin{abstract} 
    This paper presents a deterministic algorithm for resolving ambiguity in parse trees using a global mutable context. The proposed method applies a tournament-style selection process to competing derivations at each non-terminal, systematically discarding alternatives whose non-terminal subtrees are not selected by the contextual decision mechanism.
    
    Unlike approaches that rely on post-processing, the algorithm maintains semantic state throughout incremental Abstract Syntax Tree (AST) building, allowing earlier decisions to influence the resolution of future ambiguities. This context-aware strategy enables consistent and procedural disambiguation after parsing.

    It features syntactic disambiguation based on a document environment instead of relying on ad-hoc rules, thus able to model complex relationships between previous constructs and the different derivations for a same non-terminal.
\end{abstract}

\maketitle

\section{Introduction}

\subsection{Goal}
This paper aims to introduce a contextual ambiguity solver algorithm\cite{cormen2009} that clears ambiguity based on a shared mutable context.

This mechanism is a generalization of top-down, left-to-right, single-pass, context-dependent disambiguation; that can be used in grammars where ambiguity is bounded but cannot be resolved locally.

To our knowledge, no previous work explicitly formulates a deterministic, incremental, bounded-ambiguity, lexicographic elimination procedure over SPPFs with formal guarantees on termination and complexity.

\subsection{Problem}

Programming language grammars often depend on contextual information in order to resolve the meaning of certain constructs. In those cases, deciding between two or more possible derivations is only possible with imported definitions from elsewhere in the document. Many languages, like C with typedef disambiguation or Rust's function call versus tuple struct have devised clever methods or multi-stage intermediate representations in order to peek into the document context and decide for a correct interpretation of the ambiguous construct.

Throughout this paper, we'll use a classic example of ambiguity that depends on an environment to be resolved to clarify some explanations. We'll be working with a math-like domain-specific-language, where we must distinguish between implicit multiplication of a variable with a nested expression and a function call:

\begin{verbatim}
    f(0)
\end{verbatim}

The reason this is an ambiguous token sequence is that it can represent two different AST nodes and there is not enough information locally to decide.

\subsection{Limitations of Current Approaches}

We distinguish three current approaches that try to overcome this obstacle:\begin{itemize}
    \item Multi-stage AST building
    \item Pre-selected choices
    \item Name-resolution
\end{itemize}

Staged AST building with intermediate representations allows for multiple disambiguation steps before the final tree is built. Ambiguity does not exist in the grammar, but is rather about the meaning of it. Instead of having conflicting constructs in the grammar, this type of languages use nodes that represent that ambiguity without acknowledging it is the definite shape the AST will take. Its main advantage is that it allows for full contextual parsing (not just left-to-right), but requires multiple passes and ASTs, which increases memory usage and compilation time.

Pre-selected choices in certain types of parsers (like Parsing Expression Grammar parsers) solve this issue by always selecting one choice over the other. This is inherently not what we need since pre-selection replaces context.

Name-resolution is another technique, generally used by multi-stage AST builders, but it is often strictly post-syntactic, defying the purpose of using name-resolution for disambiguation in order to build the first and only AST.

\subsection{Core Idea}
\begin{figure}[h]
    \centering
    \includegraphics[width=0.5\textwidth]{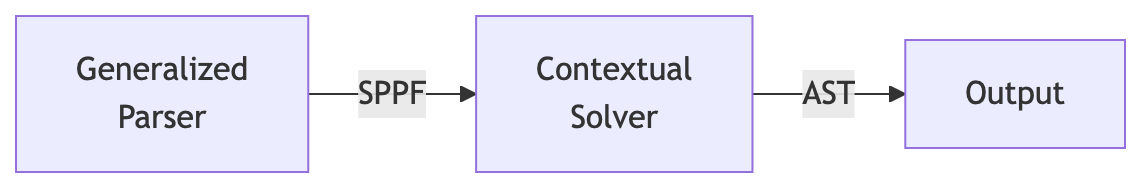}
    \caption{Two-stage parsing pipeline.}
    \label{fig:sample}
\end{figure}

Parsing\cite{aho2006compilers} becomes a two-stage pipeline, as shown in figure \ref{fig:sample}.

The generalized parser's role is to preserve all valid derivations that can be built with the supplied grammar, without deciding the final tree and return a compact backpointer representation to the solver (like a Shared Packed Parse Forest).

On the other hand, the contextual solver will compare backpointer trees at earliest points of divergence, choosing one among all candidate derivations, and producing the AST incrementally.

The solver context, which will be used to disambiguate different non-terminals, is left for host languages to define. It may include a full type analysis or be something much simpler.

\subsection{Contributions}
\begin{enumerate}
    \item A deterministic lexicographic disambiguation algorithm over SPPFs.
    \item A bounded-ambiguity complexity guarantee.
    \item A left-to-right context mutation model.
    \item Proof of termination.
    \item An implementation and empirical evaluation.
    \item A case study over a non-trivial Domain-Specific Language.
\end{enumerate}

\section{Formal Model}

\subsection{Grammar}

We define the grammar $G$ as a Context-Free Grammar\cite{sipser2006} (CFG):
$$G=(V, \Sigma, R,S)$$

where $V$ is the finite set of non-terminals, $\Sigma$ is a finite set of terminals, $S\in V$ is the start symbol, and $R$ is the finite set of production rules of the form
$$A \to \alpha \quad \text{where } A \in V \wedge \alpha \in (V \cup \Sigma)^*$$

\subsection{Uniformly Bounded Ambiguity}

We target grammars presenting localized structural ambiguity, where ambiguity is bounded per non-terminal occurrence. Most real programming languages and Domain-Specific Languages (DSLs) do not feature unbounded ambiguity, so our solver is not a general ambiguity solver for all grammars.

We define the ambiguity order $k$ as a static property of $G$:
$$\exists k \in \mathbb{N} : \forall H \in B, |H| \leq k$$

where $B$ is the finite set of backpointers (e.g. a SPPF) and $H$ is the set of valid derivations for that a non-terminal $N\in V$ over the input string $w=\Sigma^*$. Unambiguous grammars exhibit $k = 1$, and bounded ambiguity grammars have $k > 1$. It is a property of the grammar.

This assumption constraints the ambiguity per packed node, but does not bound the total number of ambiguous nodes in the forest.

For our example of \verb!f(0)!, we are dealing with a grammar with $k = 2$ since the term that holds that input can have either:\begin{itemize}
    \item two factors, \verb"f" and the nested \verb"(0)"; or
    \item one factor, \verb"f(0)", which is a function call.
\end{itemize}

we'll present our case study on section \ref{sec:6}.

\subsection{Context}

We define a mutable context $C$ which stores data defined by the host language. We assume:\begin{enumerate}
    \item All functions and procedures associated with it are deterministic.
    \item No methods of it access the compact backpointer representation ($B$).
    \item Its associated functions are guaranteed termination.
    \item Memory usage may not be constant, nor strictly linear, but may depend on how many non-terminals are appended to the AST ($t \leq n$).
\end{enumerate}

Therefore we can only claim, as we'll shortly justify, that memory usage is equal to the ambiguity order times however $C$ scales with $t$ in the individual implementation.

In our case study, $C$'s memory usage increases linearly with $t$ in the worst case, since function declarations like:

\begin{verbatim}
    f(x) := x
\end{verbatim}

can add each one entry to a set of strings that refer to variables that are functions.

When a non-terminal $A$ being built definitively to be appended to the AST, a transformation

$$ C \times A \to C_+ + N$$

which builds the non-terminal node $N$ and modifies the context is applied. This transformation mutates the context, and thus changes how disambiguation results for future (not previously processed) non-terminals.

For that reason, we say our algorithm is left-to-right. Past non-terminals can affect how future non-terminals are disambiguated but not the other way around. 

That means that \verb"f(0)" will be resolved as function call if previously we had a line specifying \verb"f(x) := x" but it will be resolved as implicit multiplication if the function declaration appears later on.

\section{The Algorithm}

\subsection{Definitions}

We define the compact backpointer representation as $B$, and $|B| = n$. The solver run function
$$R: B \times C_0 \to N_S$$

is tasked with finding the backpointer of the start non-terminal and building it into $N_S$, which is the base node of the AST. Here, $C_0$ is the default context or a context supplied by the caller.

\subsection{Inner Processing}

In order to build the start non-terminal, a build function is called. This is the first of the three core functions of our algorithm: build ($U$), solve ($Z$), and choose ($E$):\begin{enumerate}
    \item Build ($U$): $$U: H \times C \times d \to C_{(+)} + N$$
    The build function takes builds a backpointer $H$ into a non-terminal $N$. This build is definitive and mutates the context if the $d$ flag is \verb|true|, otherwise it produces exploratory builds that do not mutate the supplied $C$. The start build is always definitive.

    The $U$ function will call $Z$ in order to select the preferred derivation from $H$ according to $C$.

    \item Solve ($Z$): $$Z: H \times C \to D$$
    The solve method takes the backpointer $H$ and looks up for all derivations of it in a candidates set. Out of all the derivations, it will choose one and return it.

    In order to choose one of those candidates, these steps are performed:\begin{enumerate}
        \item We iterate over the indices of the elements of derivations.
        \item We discard those derivations that have no backpointer at this index (i.e. shorter but equal derivations).
        \item We collect all the backpointers of the derivations at this index, and we build them all with $U$ with $m$ \verb|false| into non-terminal nodes.
        \item We pop the first node from the built non-terminals and declare it the winner to start with. The final tournament winner after all the comparisons will be independent of which first node we choose, so we may select the first one\footnote{Technically, it is possible to choose a random one, from popping from a set, as this does still meet determinism.}, as strict total ordering guarantees we will arrive at the strongest node.
        \item For all the remaining built contenders, we call $E$ with the winner and the contender. If the chooser selects the winner, the winner stays in place, otherwise the contender is declared the winner.
        \item Those candidates whose backpointer at this index is the one that built the winning non-terminal node are kept in the candidates set, all others are discarded. 
        \item The index advances until there's only one candidate, and then returns it. This returned derivation is guaranteed to be the unique lexicographic maximum with respect to $\prec_C$.
    \end{enumerate}

\item Choose ($E$): $$N_1 \times N_2 \times C \to \text{either node}$$

The choose function takes two nodes $N_1$ and $N_2$ and returns a flag that indicates which node is preferred.

If the chooser reflects the intended semantic preference policy, the produced AST corresponds to the intended interpretation.

It is aided by read-only access to context, allowing for decisions based on past mutations of the context. Here, read-only methods on context may be called.

For correctness and determinism, $E$ must induce a strict total ordering over non-terminal nodes. Concretely, the strict total ordering relationship
$$
N_1 \prec_C N_2
$$

The tournament-style elimination assumes this strict total ordering; violations may yield unstable results.

This requirement may seem overly strict, or that $E$ must behave as an oracle that never fails, but in real cases it just means avoiding outputs that contradict each other (e.g. if $N_1$ is chosen instead of $N_2$ in a certain context, that the chooser called with flipped arguments yields the same result). More often than not, a problematic $E$ can just be revealed by running the same parse multiple times and comparing results.
\end{enumerate}

The solver also stores cache for exploratory-built non-terminals, that allows time complexity to drop to linear worst case instead of growing exponentially with nesting under bounded ambiguity assumption.

\section{Theoretical properties}

\subsection{Theorem: Uniqueness of Selected Derivation}

In order for determinism to hold, $\forall H \in B, Z(H,C) = $ the unique lexicographic maximum derivation under $\prec_C$. This ensures that $Z$ is a function of $H$ and $C$ and always produces the same output given the same input.

It constraints $E$ to impose a strict total ordering over the derivations.

\subsection{Strict Total Order Requirement}

Let the finite set $N = \{N_1, N_2, N_3, \dots \}$.

We define the lexicographic order induced by $\prec_C$ as a strict total order binary relation:\begin{itemize}
    \item Irreflexive: $\forall i, \neg (N_i < N_i)$
    \item Transitive: $\forall i,j,k; N_i < N_j \wedge N_j < N_k \Rightarrow N_i < N_k $
    \item Total: $\forall i, j : i \neq j; N_i < N_j \veebar N_i > N_j$
\end{itemize}

\subsection{Termination}

The algorithm is guaranteed termination since:\begin{itemize}
    \item $B$ is finite
    \item $H$ can contain at most $k$ derivations
    \item each derivation $D$ is composed of at most $r$ backpointers, therefore the iteration over indices is always guaranteed to come to an end with one candidate
    \item non-terminal node building is assumed to always terminate
    \item all procedures involving $C$ are guaranteed termination
\end{itemize}

therefore termination is guaranteed iff $B$ is a well-formed SPPF.

\subsection{Determinism}

The algorithm does not involve random choice, does not rely on race conditions, and does not involve non-deterministic guessing; so for the same input always produces the same output as long as the uniqueness theorem holds. Hence, it's deterministic.

For each input, the algorithm always follows the same finite sequence of steps and operations, thus the next operation is completely determined by the current state and input.

\subsection{Time Complexity}

Let $B$ be the compact backpointer representation and $n = |B|$. Let $k$ be the ambiguity order and $r$ the maximum length of a derivation. Let the backpointer $H \in B$.

\paragraph{Exploratory builds and memoization.}
Exploratory builds are side-effect free: they construct candidate subtrees without committing them to the AST and without mutating $C$. 
In our implementation, exploratory builds do not embed context-dependent state into the resulting node representation; contextual decisions are deferred to the chooser $E$. 
Therefore, the result of an exploratory build depends solely on the structure of $H$ and may be safely memoized.

Each backpointer $H \in B$ is thus built at most once in exploratory mode, then cached.

\paragraph{Cost per node.}
For a backpointer $H$, the solver compares $|H|$ derivations. Each derivation contains at most $r$ elements.
At each index, at most $|H|$ exploratory subtrees are constructed or retrieved from cache and compared.

Therefore, resolving a single ambiguous backpointer costs $O(|H|r)$.

\paragraph{Total cost.}
All nodes in $B$ incur this cost.

The total worst-case time complexity is therefore:

$$
O(|H|rn)
$$

since $|H| \le k$ for each individual backpointer,

$$O(knr)$$

though if we let the ambiguity density be

    $$\delta = \sum \frac{|H|}{n}, H \in B \quad \text{where }\, 1 \le \delta \le k $$
    $$\text{since } \forall H \in B, |H|\ge 1$$

then the complexity becomes

$$O(\delta nr)$$

\paragraph{Practical behavior.}
In practice, programming languages and DSLs exhibit sparse and localized ambiguity, so $\delta \ll k$. Moreover, derivations are short and typically diverge at shallow indices, causing early elimination during tournament comparison.

As a result, empirical measurements in our implementation show near-linear growth with respect to $n$ over synthetic scaling tests. This is explained by the sparsity of ambiguous nodes, early elimination during lexicographic comparison, and caching of exploratory builds.

\subsection{Memory Usage}

The memory usage is determined by:\begin{itemize}
    \item The amount of non-terminals built in the worst case for any index on $Z$: $k$.
    \item The context, for each individual candidate: $C$. How $C$ behaves asymptotically is defined by the host language.
    \item The cache for exploratory nodes grown: $n$.
    \item Each individual derivation may store at most $r$ backpointers.
\end{itemize}

Thus, memory usage scales with $n$ as $O(n+kC+kr)$.

\section{Related Work}

\subsection{Generalized Parsing}

Generalized parsers are algorithms that take an input and build a compact representation of all syntactically valid derivations.

Our two stage pipeline depends on the first step having a generalized parser that preserves all valid derivations for the non-terminals of an input so that the contextual solver can choose which one is the right one according to context and constraints.

Among generalized parsers, we find Earley\cite{10.1145/362007.362035}, GLL\cite{SCOTT20131828}, or GLR\cite{Lavie1993GLRA, CC-2004-McPeakN, article} parsers.

Our solver does not interfere with parsing, but rather chooses which of all the possible constructs suits the input according to the document state and discards the others. For that reason, we need all possible derivations preserved with a generalized parser.

\subsection{Parse Forests}

Parse forests store the full semantic information of a parse. They are produced by generalized parsers, as they allow to express ambiguity. Each parse forest has at least one parse tree, though some variants like Shared Packed Parse Forests\cite{billot-lang-1989-structure, SCOTT201055} (SPPFs) can encode exponentially many trees in polynomial memory.

For these reasons, our algorithm takes a SPPF as input, and builds the final AST incrementally instead of letting ambiguity explode.

\subsection{Semantic Predicates}

Semantic predicates are a mechanism used in some parser generators such as ANTLR\cite{10.1145/1993498.1993548, Parr1995ANTLRAP} or PEG parsers in which a boolean guard is attached to a production alternative during parsing to indicate whether that alternative is valid according to a shared context $C$.

A rule using a semantic predicate is of the form:

$$A\to \{\phi(C)\} \; \alpha$$

where $\phi$ is a function that takes $C$ and returns a boolean. The parser computes the condition at the decision point to determine whether that alternative is viable.

Our algorithm exhibits six main differences in comparison to semantic predicates:\begin{enumerate}
    \item Decision point: semantic predicates are evaluated during parsing, and if the predicate succeeds, the parser commits that derivation as the only valid one. Our solver, however, evaluates candidates after the full ambiguous parse has finished but before AST materialization.
    \item Disambiguation level: semantic predicates extend the grammar behavior with contextual conditions, effectively moving the grammar outside the pure CFG category by encoding procedural logic into productions. On the other hand, our grammar remains fully CFG, while ambiguity is preserved in the SPPF and contextual logic is isolated on the solver. We maintain a clean boundary between grammar and the derivation selection procedure, whilst semantic predicates combines them.
    \item Selection mechanism: semantic predicates operate as simple boolean filters, the alternative whose predicate succeeds is selected and competing alternatives are not materialized. Instead, our algorithm performs a tournament-style comparison, building complete derivation subtrees and being compared pairwise, eliminating candidates based on divergence with respect to the winner. It differs in that it acts as a relative preference ordering, assuming associativity and transitivity for $E$.
    \item Ambiguity preservation: semantic predicates do not preserve ambiguity, and often the first valid alternative gets chosen. In our algorithm, we preserve all valid meanings with a generalized parser and then prune candidates deterministically afterwards. While semantic predicates guide parsing, our solver interprets the parse forest according to the context.
   \item Ordering power: semantic predicates often depend on immediate semantic properties, lookahead, and symbol tables. We take a different approach by comparing the non-terminals of built subtrees supporting the lexicographic comparison accross derivations.
   \item Order of evaluation: semantic predicates are evaluated during recursive descent, allowing for context mutation as parsing advances. In our model, the context is mutated only when a node is definitively appended, while exploratory builds do not mutate the context. Disambiguation becomes strictly left-to-right at the AST construction level.
\end{enumerate}

\subsection{Attribute Grammars}

Attribute grammars\cite{10.1007/3-540-53101-7_1} are a way to associate semantic information with context-free grammars. In them, each non-terminal carries attributes and productions define equations for calculating those attributes (which may be synthesized or inherited).

The key difference resides in that an attribute grammar decorates a chosen tree, whilst our solver chooses the tree itself.

We distinguish four other differences:\begin{enumerate}
    \item Ambiguity: attribute grammars generally operate on unambiguous CFGs, and even when ambiguity is allowed the attribute system can't resolve it. Our approach requires an SPPF, treating ambiguity as expected, bounded, and resolvable instead of an error.
    \item Paradigm: attribute grammars are declarative and side-effect free. Our model uses a mutable context in which later decisions depend on earlier mutations and context evolves as nodes are committed to the AST.
    \item Evaluation strategy: attribute grammars require well-defined evaluation order, fixed dependency graphs and (often) acyclicity constraints. We, in contrast, evaluate candidates pairwise discarding losing derivations and then commit that derivation mutating context after selection.
    \item Expressiveness: attribute grammars can compute contextual properties, propagate type environments or perform name resolution; but they are not defined as selection mechanisms over ambiguous forests, or define an ordering over derivations.
\end{enumerate}

\subsection{Ordered Choice in PEGs}

In Parsing Expression Grammars\cite{10.1145/982962.964011,Medeiros_2014}, there exists a choice (\verb"/") operator which performs a deterministic, priority-based selection. For rules like:

$$A \to d_1 / d_2$$

it allows the parser to have $d_2$ as fallback in case $d_1$ didn't match.

PEG priority is static, defined at grammar level, and only depends on the local success of the leftmost derivation. Our solver is dynamic, depends on a mutable context, selects one derivation among all valid derivations, compares subtrees structurally and prunes candidates progressively.

We observe four other differences:\begin{enumerate}
    \item Context: PEG ordered choice is locally greedy, selecting $d_2$ iff $d_1$ fails. Our algorithm is globally comparative and sees all valid derivations simultaneously and commits one.
    \item Ambiguity handling: PEGs do not represent ambiguity and output exactly one valid parse, while ambiguity is encoded away via rule ordering. Our approach relies on a generalized parser preserving all meanings compactly and treating ambiguity as bounded and intentional.
    \item Expressiveness: PEG ordered choice alone cannot express preference based on symbol tables (doing so requires embedding semantic predicates), previously chosen derivations, or structural comparison between two successful parses. Our solver's decision can be influenced by earlier AST commits, deep subtree comparison and context-sensitive ordering.
    \item Grammar: PEGs have disambiguation encoded into rule order so reordering derivations changes semantics, since grammar and disambiguation are intertwined. In our method the grammar remains a pure CFG, disambiguation logic is externalized into the solver, and the grammar can remain declarative and ambiguity-preserving.
\end{enumerate}

\subsection{SDF Filters}

SDF filters\cite{m.g.j._2001} allow for declarative disambiguation over a parse forest. They also require a generalized parser, but disambiguation is resolved applying multiple filters to the forest and then materialize the remaining tree.

Classical SDF filters are declarative and grammar-level, and do not model mutable document state that evolves during disambiguation. These filters encode priority, associativity, restrictions, constraints and preferences; so the rules are applied sequentially and then the AST is built.

The main difference between our algorithm and SDF filters is that our solver performs context-sensitive comparison between subtrees based on a mutable context. Our model allows for mutable context that disambiguates future non-terminals based on how earlier ones modified the context.

The design philosophy differs in that our solver produces the start non-terminal when run:
    $$R:B\times C_0 \to N_S$$

whilst SDF defines:

    $$\mathcal{F}: \text{SPPF} \to \text{SPPF} $$

\subsection{Summary}

Table~\ref{tab:comparison} summarizes the main differences between our solver and related disambiguation mechanisms.

\begin{table*}[h]
\centering
\small
\begin{tabularx}{\textwidth}{l *{5}{>{\centering\arraybackslash}X}}
\textbf{Property} & \textbf{Our Solver} & \textbf{Semantic Predicates} & \textbf{Attribute Grammars} & \textbf{PEG Ordered Choice} & \textbf{SDF Filters} \\
\hline
Requires generalized parser & Yes & No & No & No & Yes \\
Preserves ambiguity explicitly & Yes (SPPF) & No & Typically No & No & Yes \\
Grammar remains pure CFG & Yes & No & Yes & No (priority-based) & Yes \\
Disambiguation location & Post-parse & During parsing & After tree selection & During parsing & Post-parse \\
Context mutability & Yes (left-to-right) & Yes & No (declarative) & No (static priority) & No (declarative) \\
Context affects future decisions & Yes & Yes & Limited & No & No \\
Pairwise structural comparison & Yes & No & No & No & No \\
Deterministic outcome & Yes & Yes & Yes & Yes & Certain configurations \\
Bounded ambiguity required & Yes & No & No & No & No \\
Separation of grammar and disambiguation & Strict & Mixed & Strict & Mixed & Strict \\
\end{tabularx}
\caption{Comparison between environment-sensitive lexicographic disambiguation and related techniques.}
\label{tab:comparison}
\end{table*}

\section{Case Study: A Mathematical DSL}
\label{sec:6}

\subsection{Grammar}

The grammar of our DSL has 24 non-terminals and 25 token types. Our language has only one type of ambiguity, with $k = 2$. See the full grammar specification for our language in appendix \ref{app:A}.

\subsection{Ambiguity}

\begin{figure}[b]
    \centering
    \includegraphics[width=1\linewidth]{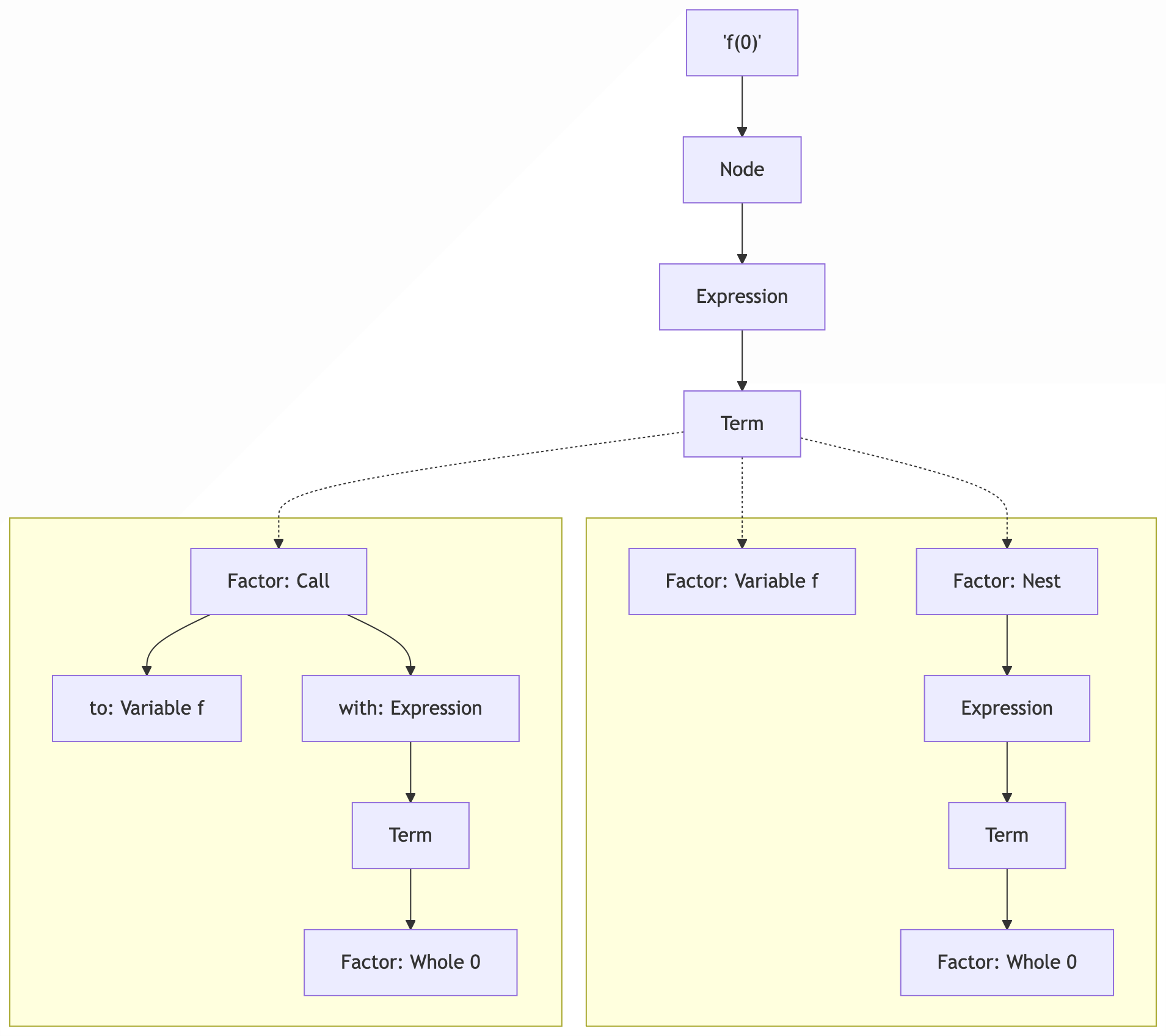}
    \caption{Parse subtree for function call (simplified).}
    \label{fig:placeholder}
\end{figure}

Ambiguity arises in the programs of our language when we find structures like\begin{itemize}
    \item \verb"f()"; or
    \item \verb"f(0)"
\end{itemize}

because it can be both interpreted as a term with two factors (\verb"f" and \verb"(0)") or a function call, as show in figure \ref{fig:placeholder}.

So we've implemented a $C$ that stores a set of strings for variables that have been declared as functions $C_f$, and a $E$ that disambiguates between between a call or the first factor of the two-factor sequence by looking up whether $v \in C_f$ where $v$ is the variable string.

Subsequently, we have also defined that when a \verb!Function! AST node is built, it inserts its variable into $C_f$.

We will now walk through two examples to show the solver in action.

\paragraph{Implicit multiplication}
For the full document:
\begin{verbatim}
    x(x + 1)
\end{verbatim}

Our solver finds the earliest point of divergence at index zero of the term, where the different derivations produce different exploratorily-built nodes:\begin{itemize}
    \item A factor containing a function call to function \verb|x| with the expression of the first argument being \verb|x + 1|.
    \item A factor containing a raw variable, \verb|x|, as the first factor of the sequence.
\end{itemize}

Therefore, it looks up in $C_f$ whether \verb|x| was declared as a function. Since it wasn't declared as a function before, $E$ deliberately selects the raw variable factor as the correct interpretation.

\paragraph{Function call}
Suppose we have the document:
\begin{verbatim}
    f(x) := x
    f(0)
\end{verbatim}

First, the solver will find the unambiguous \verb|Function| node. Thus, when it is committed to the AST, it inserts into $C_f$ that \verb|f| was declared as a function.

Later, when $E$ finds itself disambiguating \verb|f(0)|, it will look up $C_f$ and see \verb|f| was declared as a function, therefore choosing the interpretation that \verb|f(0)| is a function call.

As we see in this example, the strict total ordering requirement is easily imposed, and no matter which winner was selected at first by the algorithm, we'll arrive at function call iff the variable was declared as a function.

The strict ordering requirement is not an oracle; it corresponds to defining a deterministic precedence policy over competing interpretations. In practice, this is simpler than defining consistent semantic predicates across the grammar.

\subsection{Nested Ambiguity}

Memoization is essential so pathological input like:
\begin{verbatim}
    f(f(f(f(f(f(f(f(f(f(f(0)))))))))))
\end{verbatim}

can be processed without collapsing into exponential time. In this example, $k$ is still equal to 2 because the nested expressions are not ambiguous themselves but rather their factors. However, caching is necessary to process it in polynomial time. Our implementation is able to solve a few hundred nested levels with memoization, avoiding exponential explosion.

\subsection{Implementation and Evaluation}

\paragraph{Implementation.}
The solver and generalized parser were implemented in Rust 1.95-nightly. 
We built a custom Earley parser producing an SPPF and a standalone contextual solver.

The forest is roughly represented as:

\begin{verbatim}
HashMap<Backpointer, HashSet<Vec<Backpointer>>>
\end{verbatim}

where each key corresponds to a non-terminal span and maps to its set of derivations. 
The grammar has maximum derivation length $r = 8$.

Since exploratory builds are side-effect free, each backpointer is built at most once.

The context is merely a simple struct:

\begin{verbatim}
struct Context { functions: HashSet<String> }
\end{verbatim}

It stores declared functions and mutates only when a \verb|Function| node is definitively appended. 
Exploratory builds use cloned contexts.

The DSL defines 24 AST node types, built incrementally and heap-allocated.

\paragraph{Experimental setup.}
Experiments were run on a MacBook Air M4 (10-core CPU, 16GB RAM, macOS) compiled in release mode with full LTO and \verb|opt-level = 3|. 
Each measurement was repeated five times.

\paragraph{Results.}
We measured solver runtime as a function of SPPF size $n$ for both unambiguous and ambiguous inputs.

Unambiguous input:

\begin{center}
\begin{tabular}{r r}
$n$ & mean time (ms) \\
\hline
5k   & 0.23 \\
50k  & 1.66 \\
175k & 5.07 \\
657k & 13.01 \\
2.5M & 33.45 \\
\end{tabular}
\end{center}

Ambiguous input:

\begin{center}
\begin{tabular}{r r}
$n$ & mean time (ms) \\
\hline
4k   & 0.58 \\
36k  & 3.16 \\
124k & 7.75 \\
453k & 17.06 \\
1.7M & 45.35 \\
\end{tabular}
\end{center}

Observed growth is close to $\Theta(n^{0.7})$ in both cases. 
Ambiguity seems to increase constant factors but not change the exponent.
Measured ambiguity density was $\delta \le 1.024$, and no pathological slowdowns were observed, including nested ambiguous inputs.

The current implementation uses recursive construction and may exhibit stack overflow on artificially deep nesting (i.e. $>500$ depth levels). This does not affect the theoretical guarantees but suggests that an iterative formulation would be preferable for production-scale inputs.

These results empirically support the $O(\delta n r)$ complexity under bounded ambiguity.

\section{Discussion}

\subsection{Limitations}

Although our proposed solver provides a deterministic and context-sensitive mechanism able to resolve bounded ambiguity, it is subject to several important limitations.

\paragraph{Bounded Ambiguity Requirement.}
The linear-time guarantee depends on $k$ being constant. Grammars in which a single non-terminal may admit an arbitrary number of derivations over the same span violate this assumption and therefore may cause the solver to explode combinatorily. This approach is inherently unsuitable for grammars without structural constraints.

\paragraph{Left-to-right commitment.} Context mutation happens strictly left-to-right as it's the order in which the AST is built. Earlier decisions might influence later ones, but not vice versa. Ambiguities whose resolution depends on future constructs might not be resolved as intended without extending the model with rollback.

\paragraph{Dependence on a well-behaved chooser}
Correctness and determinism rely on the chooser function $E$ defining a strict total ordering. In practice, it must behave transitively and coherently across comparisons. If $E$ violates this guarantee, the tournament-style pruning process may yield unstable results. The framework does not enforce ordering laws formally, and they must be guaranteed by the host implementation.

\paragraph{Exploratory build cost.}
Although caching exploratory builds mitigates costs, exploratory subtree construction may introduce significant constant-factor overhead on deep and ambiguous nested constructs even when asymptotic bounds remain linear.

\paragraph{Context predictability.}
Because the context $C$ is mutable and host-defined, its expressiveness is unconstrained. It thus introduces the risk of large complex increasing memory usage beyond practical limits, or subtle context mutations making disambiguation behavior harder to reason about.

\subsection{Applicability}

The solver is best suited for languages where ambiguity arises from well-structured syntactic overlap dependent on evolving document state rather than global unstructured ambiguity. Its design favors environments targeting a final AST to be produced deterministically, with the aid of contextual information to guide syntactic interpretation.

\paragraph{Programming languages with context-dependent syntax.}
Many modern programming languages exhibit localized ambiguities whose resolution depends on symbol tables, declarations, or type environments (e.g. distinguishing variable references from type names, or function calls from constructors). The solver provides a principled alternative to ad-hoc grammar refactoring or multi-phase AST building, allowing those languages to retain a clean CFG while delegating contextual choice to the solver.

\paragraph{Domain-specific languages.}
Domain-Specific Languages (DSLs) often introduce environment-sensitive constructs or user defined operators, amongst other self-defined behavior. Because ambiguity in such DSLs is typically small, bounded, and follows coherent rules, the solver is a practical fit.

\paragraph{Mathematical and symbolic notations.}
Languages modeling mathematical notation frequently contain ambiguities between implicit multiplication, function application, and parenthesized grouping. These ambiguities are rarely resolvable locally but depend on previously introduced declarations. The left-to-right context mutation model aligns naturally with such semantics.

Conversely, the approach is less appropriate for languages with pervasive or unbounded ambiguity, for specifications requiring fully declarative disambiguation, or for systems that depend on global whole-program inference before syntax can be interpreted. Within its intended design space (bounded, contextual ambiguity), the solver provides a structured and deterministic resolution mechanism that integrates syntactic and contextual reasoning without sacrificing grammar clarity.

\bibliographystyle{acm}
\bibliography{references}

\appendix

\section{Case Study Grammar}
\label{app:A}

Below is the full Extended Backus-Naur Form (EBNF) specification for our language. This grammar is converted to BNF form automatically and then parsed, but for simplicity we will only provide the EBNF form.

The grammar is unambiguous except for the call / implicit multiplication overlap.

\begin{Verbatim}[breaklines=true]
//> EBNF -> START
Start -> (NEWLINES? Level1 (NEWLINES Level1 SPACES?)*)? NEWLINES? ENDOFFILE

//> EBNF -> 1ºLEVEL
Definition -> Variable DEFINITION Level2
Function -> Variable OPEN (Variable (COMMA Variable)* SPACES?)? CLOSE DEFINITION Level2
Node -> Level2
Equation -> Level2 EQUALITY Level2
Use -> USE MODULE

//> EBNF -> 2ºLEVEL
Expression -> (SIGN SPACES?)* Level3 ((SIGN)+ Level3)*

//> EBNF -> 3ºLEVEL
Term -> Level4 ((OPERATOR)? Level4)*

//> EBNF -> 4ºLEVEL
Factor -> Level5 (EXPONENTIATION Level2 EXPONENTIATION)?
Limit -> LIMIT Variable TO Level2 SIGN? OF Nest (EXPONENTIATION Level2 EXPONENTIATION)?

//> EBNF -> 5ºLEVEL
Infinite -> INFINITE
Variable -> IDENTIFIER
Nest -> OPEN Level2? CLOSE
Tensor -> ENTER (Level2 (COMMA Level2)* SPACES?)? EXIT
Whole -> NUMBER
Absolute -> PIPE Level2 PIPE
Undefined -> UNDEFINED
Rational -> RATIONAL
Call -> Variable OPEN (Level2 (COMMA Level2)* SPACES?)? CLOSE

//> EBNF -> LEVELS
Level1 -> Definition | Function | Node | Equation | Use
Level2 -> Expression
Level3 -> Term
Level4 -> Factor | Limit
Level5 -> Infinite | Variable | Nest | Tensor | Whole | Absolute | Undefined | Rational | Call
\end{Verbatim}

\end{document}